\documentclass[lettersize,journal]{IEEEtran}
\usepackage{amsmath,amsfonts}
\usepackage{algorithmic}
\usepackage{algorithm}
\usepackage{array}
\usepackage[caption=false,font=normalsize,labelfont=sf,textfont=sf]{subfig}
\usepackage{textcomp}
\usepackage{stfloats}
\usepackage{url}
\usepackage{verbatim}
\usepackage{graphicx}
\usepackage{cite}
\def\BibTeX{{\rm B\kern-.05em{\sc i\kern-.025em b}\kern-.08em
    T\kern-.1667em\lower.7ex\hbox{E}\kern-.125emX}}

\usepackage{booktabs}
\usepackage{hyperref}

\hypersetup{
    colorlinks=true,
    linkcolor=blue,
    filecolor=magenta,      
    urlcolor=blue,
}

\begin{document}

\def\P910{ITU-T Rec.~P.910}

\title{A crowdsourcing approach to video quality assessment}
\author{Babak Naderi, Ross Cutler} 


\maketitle

\begin{abstract}
The gold standard for measuring video quality is the subjective test, and the most prevalent is the \P910, a lab-based subjective standard in use for the past two decades. However, in practice using \P910 is slow, expensive, and requires a lab, which all create barriers to usage. As a result, most research papers that need to measure video quality don't use \P910 but rather metrics that are not well correlated to subjective opinion. We provide an open-source extension of \P910 based on crowdsourcing principles which address the speed, usage cost, and barrier to usage issues. We implement Absolute Category Rating (ACR), ACR with hidden reference (ACR-HR), Degradation Category Rating (DCR), and  Comparison Category Rating (CCR). It includes rater, environment, hardware, and network qualifications, as well as gold and trapping questions to ensure quality. We have validated that the implementation is both accurate and highly reproducible compared to existing \P910 lab studies. 
\end{abstract}


\section{Introduction}
\IEEEPARstart{M}{easuring} the quality of video is an important task in many engineering areas, such as video codec development, video enhancement, and video telecommunication systems. As a result, there are many standards and metrics for measuring video quality, though the gold standard is the subjective test done in a controlled lab environment. The most prevalent of these standardized subjective tests is the \P910 \cite{noauthor_itu-t_2021}. However, using \P910 in practice is slow due to the recruitment of test subjects and the limited number of test subjects, and expensive due to paying qualified test subjects and the cost of the test lab. The speed and cost result in the vast majority of research papers not using \P910 but rather objective functions that are not well correlated to subjective opinion. 

An alternative to lab-based subjective tests is to crowdsource the testing, and there are several such systems that do this (see Table \ref{tab:comparison}). However, none of these identified systems have the rater, environment conditions, and hardware qualifications that \P910 requires. In addition, these existing systems have not been rigorously validated to show they are accurate compared to a \P910 lab-based study and give reproducible results. We provide a crowdsourced implementation that includes rater, environment, hardware, and network qualifications, as well as gold and trapping questions to ensure quality. We include a validation study that shows it is both accurate and highly reproducible compared to existing \P910 lab studies. The tool is open-sourced and can be used on the Amazon Mechanical Turk platform for wide-scale usage. We recently used the tool in the CVPR 2022 CLIC challenge\footnote{\url{http://compression.cc/}} to provide the challenge metric for machine learning-based video codecs.

A significant advantage of the crowdsourcing approach to labeling data is its scalability due to low cost,  availability of workers, and speed. Crowdsourcing has been widely used to label very large image datasets like ImageNet \cite{deng_imagenet:_2009}, as well as to label millions of audio clips for speech quality assessment for use in academic challenges and speech enhancement development \cite{naderi_open_2020,cutler_crowdsourcing_2020,naderi_subjective_2021,dubey_icassp_2022,cutler_icassp_2022,reddy_dnsmos_2021,reddy_dnsmos_2022,purin_aecmos_2022,diener_interspeech_2022}. This crowdsourcing tool enables millions of video clips to be accurately, inexpensively, and quickly labeled for video quality assessment, which is necessary data to develop machine learning-based video codecs, video quality enhancement, and to optimize video-based telecommunication systems.

The vast majority of published video compression research papers provide results using the objective metrics Peak signal-to-noise ratio (PSNR) \cite{gonzalez_digital_2006}, structural similarity index measure (SSIM) \cite{wang_image_2004}, multi-scale SSIM (MS-SSIM) \cite{wang_multiscale_2003}, and Video Multi-Method Assessment Fusion (VMAF) \cite{li_toward_2016}, e.g., \cite{lin_m-lvc_2020, hu_fvc_2021, rippel_elf-vc_2021}. However, it has been shown that PSNR, SSIM, and MS-SSIM are not well correlated to subjective opinion \cite{li_toward_2016, seshadrinathan_subjective_2010}. VMAF has been trained on traditional DSP-based video codecs such as H.264/AVC \cite{wiegand_overview_2003}, but has not been shown to correlate with the newer deep learning-based codecs. Moreover, VMAF hasn't been trained for longer-term effects like recency, primacy, and rebuffering \cite{li_vmaf_2018}. Therefore for video codec research, there is no alternative to subjective tests to evaluate and compare codecs at this time.

Section \ref{sec:related_work} provides an overview of related work, Section \ref{sec:implementation} describes our tool's implementation, Section \ref{sec:validation} gives the tool's validation, and Section \ref{sec:conclusions} gives conclusions. 

\section{Related work}
\label{sec:related_work}
A recent review of subjective quality assessment tools is given in \cite{testolina_review_2021}. \P910 \cite{noauthor_itu-t_2021} provides a general subjective video quality assessment standard for multimedia applications. \P910 includes ACR, ACR-HR, DCR, and paired comparison (PC) methods, as well as rater qualifications, environment conditions, and video playback procedures. ITU-T Rec.~P.911 is a counterpart of \P910 but for audiovisual signals. ITU-T Rec.~P.912 \cite{noauthor_recommendation_2016} provides a target-specific subjective video quality assessment standard, such as for faces, license plates, etc. ITU-T Rec.~P.913 \cite{ITU-P913} considers different displays and testing environments and provides flexibility on the rating scale and modality with mandatory reporting of test requirements \cite{pinson2014new}. ITU-T Rec.~P.918 \cite{ITU-P918} details subjective assessment methods for five perceptual video quality dimensions, which can provide diagnostic information on the source of observed degradation. Finally, ITU-R BT.500 \cite{ITU-BT500} focuses on the video quality of broadcast television signals in a highly controlled environment.

Tominaga et al.~\cite{tominaga_performance_2010} conducted a comparison of eight different subjective video quality assessment methods and found that ACR was the most suitable for statistical reliability, assessment time, and ease of evaluation. 

Keimel et al.~\cite{keimel_qualitycrowd-framework_2012} describes an open-source tool QualityCrowd that supports ACR video quality assessment. QualityCrowd is extended by \cite{upenik_large-scale_2021} to include a Double Stimulus Continuous Quality Scale (DSCQS). Rainer et al.~\cite{rainer_web_2013} describe the tool WESP, an open-source tool that supports ACR, ACR-HR, DCR, and PC. 

Jung et al.~\cite{jung_isoiec_2021} provide a methodology to conduct remote subjective video quality assessment studies in which the raters download videos and view them manually on their devices. While the methodology suggests the rater checks visual acuity and color blindness, it doesn't provide any such tests. There are also no tests for environmental conditions or hardware setup, which we will show is essential. 

\begin{table}[]
\begin{center}
\caption{Open-source crowdsourcing video quality assessment systems}
\label{tab:comparison}
\scalebox{0.75}{
\begin{tabular}{|c|c|c|c|c|c|c|c|}
\hline
\textbf{Tool} & \textbf{Measures} & \textbf{Rater} & \textbf{Viewing} & \textbf{HW} & \textbf{Network} & \textbf{Accur.} & \textbf{Repro.} \\
& & \textbf{qual.} & \textbf{cond.} & & & & \\
\hline
QualityCrowd & ACR, DSCQS & N & N & N & N & Y & N \\
\cite{keimel_qualitycrowd-framework_2012,upenik_large-scale_2021} & & & & & & & \\
\hline
WESP \cite{rainer_web_2013} & ACR, ACR-HR, & N & N & N & N & N & N \\
 & DCR, PC & & & & & & \\
\hline
avrateNG \cite{rao_towards_2021} & ACR & N & N & N & N & Y & N \\
\hline
Ours & ACR, ACR-HR, & Y & Y & Y & Y & Y & Y \\
 & DCR, CCR & & & & & & \\
\hline
\end{tabular}
}
\end{center}
\end{table}

\section{Implementation}
\label{sec:implementation}
Our implementation\footnote{\url{https://github.com/microsoft/P.910}} can be used either as an integrated survey in Amazon Mechanical Turk (AMT) or as an external survey deployed on a dedicated Web server\footnote{Currently the integrated surveys within AMT are not able to playback videos in full-screen mode, so we strongly recommend to deploy the test as an external survey.}. Consequently, AMT or any other crowdsourcing platform can be used for recruiting and paying test participants. 
We also provide a lightweight container-based web application that can serve the experiment. The web application can easily be deployed on any Linux virtual machine. As a result, this implementation can be used for both crowdsourcing tests or remote testing with a dedicated panel of participants.

The open-source implementation includes the Absolute Category Rating (ACR), ACR with hidden reference (ACR-HR), Degradation Category Rating (DCR), and Comparison Category Rating (CCR). All methods can be used with either a five or nine-point discrete Likert scale. We also followed and extend best practices on video and speech quality assessment in crowdsourcing \cite{hossfeld_best_2014, ITU-PSTR-CROWDS} in our implementation.

\subsection{Tools}
We provide a set of program scripts to ease the interaction with the system and avoid operation errors. The scripts are used to prepare the subjective test (e.g., create customized trapping sequences), process the submitted answers (i.e., data cleansing, aggregating ratings, and generating reports), and interact with AMT (e.g., sending bonuses to workers).

\begin{figure}[htbp]
\centerline{\includegraphics[width=\columnwidth]{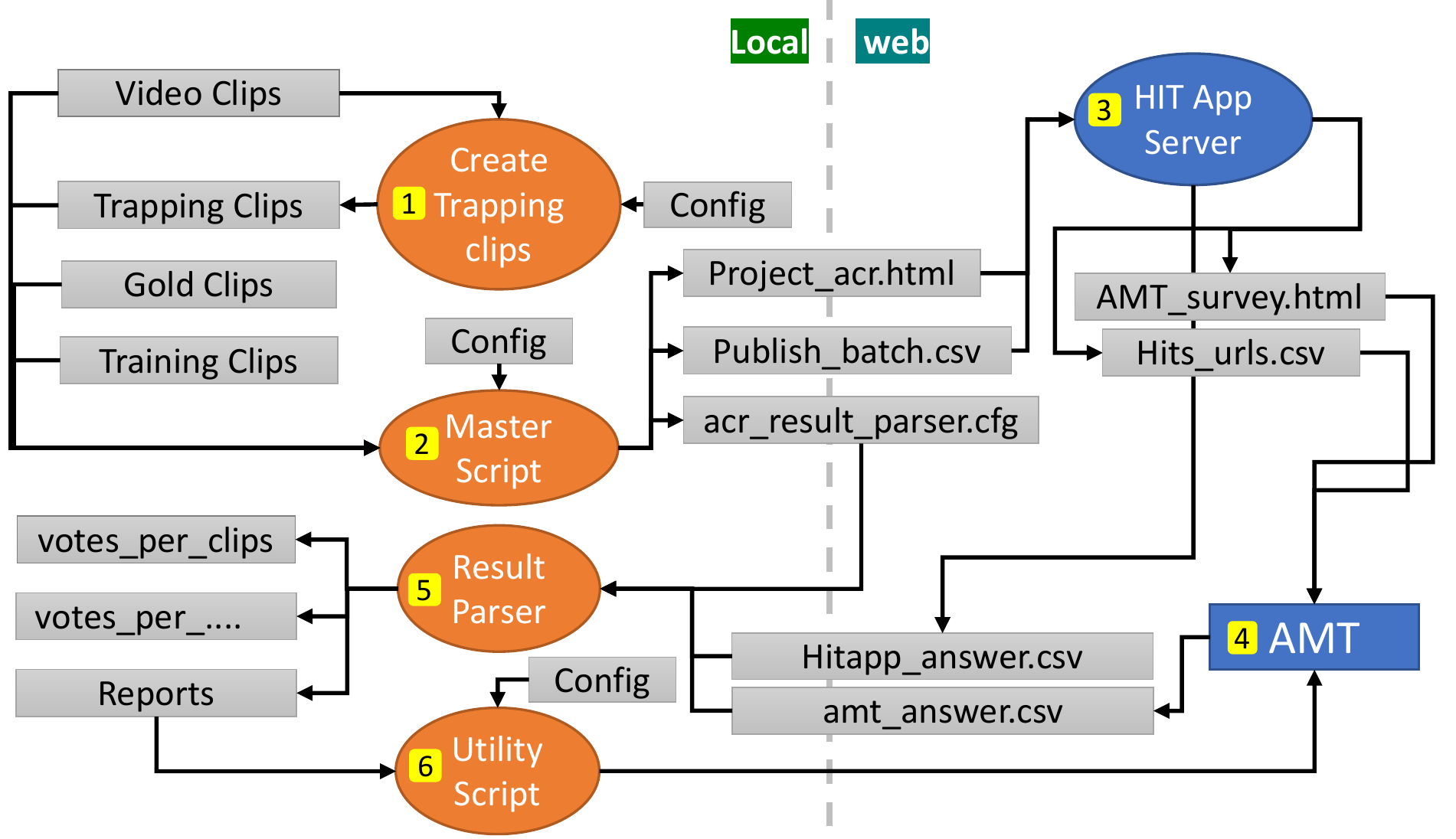}}
\caption{Data Flow Diagram. }
\label{fig_dfd}
\end{figure}

In the first step (see Figure~\ref{fig_dfd}), trapping clips customized to the dataset under the test should be created using the corresponding script. They will be created by adding a text message, in the middle of selected existing video sequences, asking participants to select a specific score when rating this trapping clip. Next, a test configuration and URLs for test sequences (in the case of DCR, CCR and ACR-HR also reference sequences), trapping clips, gold, and training clips should be provided to the master script. It creates the HTML template of the test, a list of variables (dynamic content), and a configuration file for the result parser. Gold clips are video sequences whose quality is known to the experimenter (either excellent or bad) \cite{ITU-P808}. A test can be created in the HIT App Server by providing the HTML template and a list of variables generated in the last step. Finally, a generic project description and list of URLs can be downloaded and used to create a new test in AMT (or any other crowdsourcing platform).

When the test is finished, the submitted answers to AMT and HIT App Server should be provided to the result parser script in addition to the configuration file which was created in step 2 (see Figure~\ref{fig_dfd}). It will perform the data cleansing process, and aggregate the valid and reliable ratings over the test sequence and over the test condition (i.e., Hypothetical Reference Circuits\footnote{HRC is a treatment applied to source video to create the test sequence.} - HRCs) if they exist. In addition, reports on the list of bonus assignments, and lists of the submissions to be accepted/rejected or extended will be generated.

\subsection{Test components}
\begin{figure}[htbp]
\centerline{\includegraphics[width=\columnwidth]{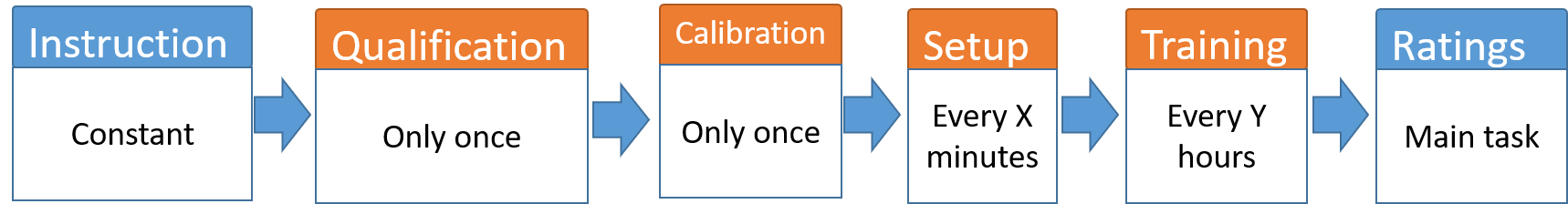}}
\caption{The crowdsourcing test from the participant's perspective.}
\label{fig_info}
\end{figure}

The subjective test is organized in different sections from the participant's perspective (see Figure~\ref{fig_info}). Each section is designed to instruct the test participant, qualify the participant, their environment, their setup, and collect their votes. The \textit{instructions} and \textit{ratings} sections are included in all tests, whereas the \textit{qualification}, and \textit{calibration} only need to be performed once per test. The \textit{setup} and \textit{training} sections are shown  (e.g., once per hour).
It is recommended to keep the crowdsourcing test session short (less than 15 minutes), therefore we present a limited set of test sequences in the rating section. We encourage participants to take part in more than one session by providing monetary incentives and dynamic sections (i.e., the next test session will be significantly shorter than the first one). 

A set of automatic measurements are performed on the test's loading time. The experimenter can restrict test participants to specific viewing devices (i.e., mobile, PC, or both), minimum screen refresh rate, and minimum resolution.

\subsubsection{Video playback}
We developed a video playback component based on HTML5. To avoid the confounding effect of the network connection, it fully downloads all the video materials on the browser's local storage. The videos are also played in full-screen mode, and the participants must watch the video until the end before casting their votes. The player records the playback duration of videos which will be used later in the data cleansing process. The experimenter can decide whether the videos should be presented in their original size or scaled to fill the participant's viewing device. In the case of the DCR or CCR test, the reference and processed video sequence will be shown in a sequence with a gray screen in between for one second.

\subsubsection{Qualification}
Within the qualification section, the eligibility of test participants is evaluated. According to the ITU-T Rec. \P910, participants should be screened for normal color vision and normal or corrected-to-normal visual acuity (i.e., no error on the 20/30 line of a standard eye chart) \cite{ITU-P910}.
The standard Ishihara test for color vision \cite{clark1924ishihara} includes 15 plates in the normal and 6 plates in the short version, which are both too long for a crowdsourcing test. In a prestudy, we invited 300 participants from AMT (34\% male) and 191 participants (91\% male) from two internet communities dealing with color vision deficiencies. Both groups participated in the full Ishihara test in which 6\% of participants from AMT and 96\% from the online forums were detected as color blind. Applying a decision tree classifier with entropy as the criterion revealed that only using Plates 3 and 4 reached 98\% accuracy (sensitivity 0.996, specificity 0.95). Consequently, we only use these two plates in the qualification section.

For the visual acuity test, we use the Landolt ring optotypes as recommended by ISO 8596:2017 \cite{iso_8596}. In this test, participants are presented with broken rings (like the letter C). The gap can be in 8 directions (every 45 degrees). The diameter of the ring is 5 times the size of the gap. The visual acuity is the inverse value of the gap size (in arc minutes) of the smallest identified Landolt ring \cite{iso_8596}. In each row (i.e., ring size) 5 samples should be presented and the participant should answer 3 or more correctly to pass that size.
Our implementation of the visual acuity test consists of two steps: 1) setup, and 2) answer to up to 5 Landolt rings on the specific size. In the setup section, the size of a pixel in the user’s screen is estimated by asking them to adapt the size of a given picture (here a credit card) on their screen by clicking on +/- buttons to reach the specified size. In addition, we asked them to sit in the range of 50 to 75 cm from the screen during the test. Although the maximum viewing distance of participants is limited to a distance in which they can easily interact with their device, we cannot enforce a minimum distance. Given the pixel size, the corresponding Landolt ring size is calculated which correctly answering them in 50 cm distance shows a minimum of 20/30 visual acuity. The participant passes the visual acuity test if they correctly select the direction of three out of the five Landolt rings. Finally, participants can only continue to the next sections if they successfully pass the qualification.

\subsubsection{Display calibration and instructions}
Participants are asked to set the resolution of their device to default or recommended value suggested by their operating system. We also ask them to perform display color calibration using methods provided by their operating system. This section provides information on how to perform these tasks for Windows and Mac operating systems and will only be shown once during the test.

In the instruction section, a sample video for some of the perceptual quality dimensions~\cite{schiffner2017defining, ITU-P918} (i.e., fragmentation, discontinuity, and uncleanness) is presented to the participants. They are also informed that impairments can happen in a specific area or time within a video clip and the impairments are not limited to the presented samples.

\subsubsection{Setup}
\label{setup}
\begin{figure}[tb]
\centerline{\includegraphics[width=0.6\columnwidth]{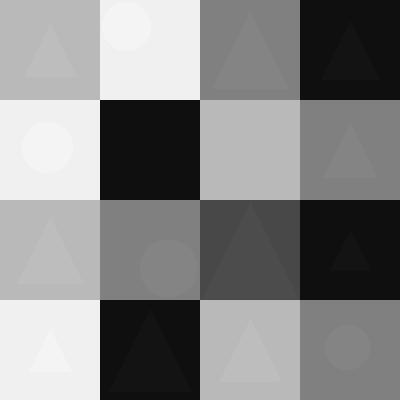}}
\caption{Matrix of squares with 4 circles and 10 triangles used for evaluating screen calibration.}
\label{fig_martix}
\end{figure}

In the setup section, the viewing conditions including the brightness of the screen, room light, and viewing distance are evaluated. The \P910 provides a detailed specification for viewing conditions that can only be measured in a controlled laboratory environment using professional devices. We created a brightness/light calibration task in which participants should count the number of geometric shapes presented in a picture. They can insert their answers and verify them. Given the correct answer, they can continue to the next steps whereas on mismatch they are encouraged to modify the brightness of their screen, lighting of the environment, or repeat the display calibration task until they can insert the correct number for each picture. The presented picture is a matrix of 4x4 squares. Each square has a different level of gray background and may contain a triangle or circle in different sizes and locations. The foreground color of the shapes is close to the background color of the square. A pretest, with a limited group of experts, showed that the calibration task performs best when the difference between foreground and background colors is four integer points in RGB24 color space. Figure~\ref{fig_martix} illustrates an example matrix picture used in our tests. Within the setup section, a second matrix is also presented for which no feedback is given. In post-processing, submissions will be filtered based on their answers to the second matrix question.
It should be noted that to provide online filtering and feedback, encrypted correct answers to some of the tests are included in the HTML/JS code and a limited number of re-tries are provided in case of failure. All the qualifications are also assessed separately in the post-processing step.

\begin{figure}[tb]
\centerline{\includegraphics[width=0.8\columnwidth]{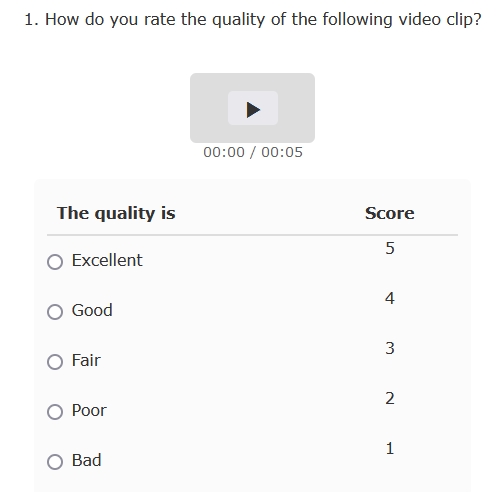}}
\caption{Video playback control and rating scale implemented for ACR test.}
\label{fig_acr}
\end{figure}

\begin{figure*}[htb]
\centerline{\includegraphics[width=1\textwidth]{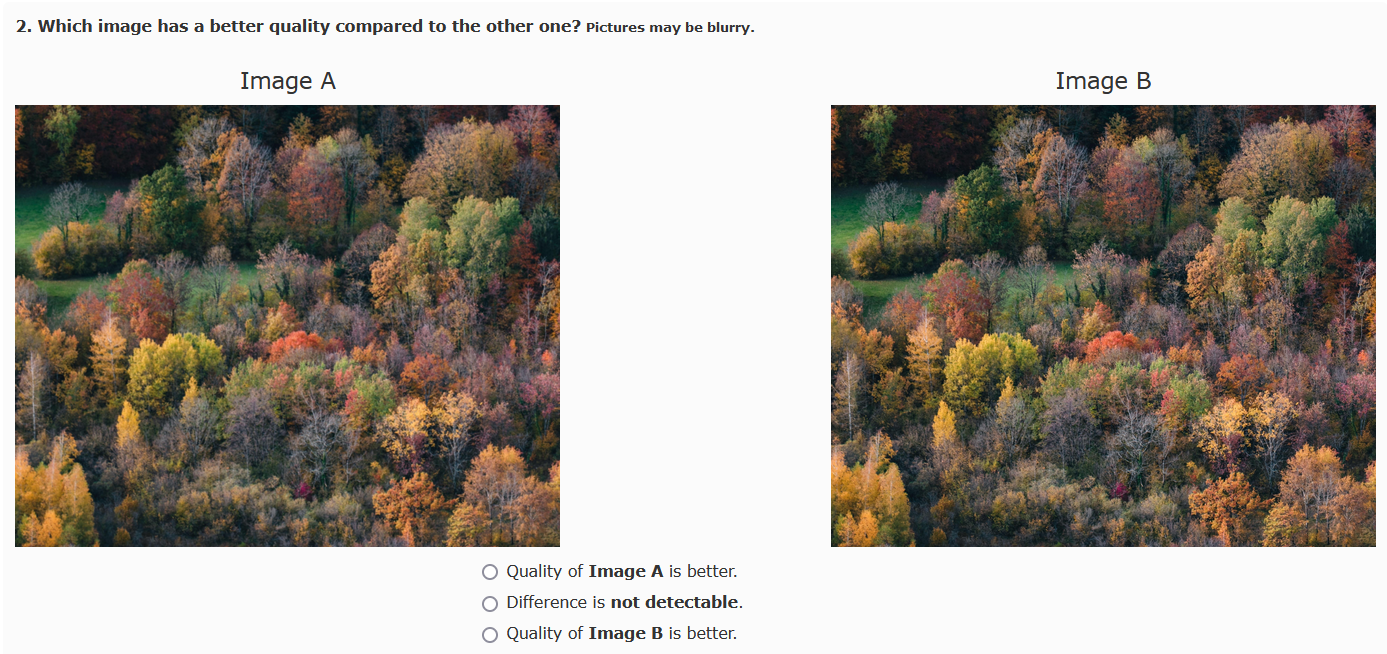}}
\caption{One of the questions used in a paired-comparison test to check viewing distance. Participants recognize the difference when they are too close to their display.}
\label{fig_distance}
\end{figure*}

The crowdsourcing test also includes a short paired-comparison task to evaluate participants' viewing distance. The task is inspired by \cite{naderi2020application} which is the recommended method to evaluate crowd workers' setup and environment in a speech quality assessment test \cite{ITU-P808}.
In this task, three pairs of images are presented to the participants who are asked to sit at a distance of 1.5 times their screen height. In each pair, one of the images is distorted with a blur effect. Participants are asked to select the image of better quality. They can choose between either of the images or select that both have the same quality. In a separate experiment, we used blur effects with different radiuses from 1 pixel to 9. A small group of participants took part in six Just-Noticeable Difference (JND) quality tests (adaptive staircase method 2 down–1 up)~\cite{treutwein1995adaptive} with two different screen sizes and three different viewing distances (too close, expected, too far). Using the results, we selected three pairs for this task: 1) the difference is mostly recognized if the participant is too close to the screen, 2) the difference is correctly recognized if the participant is seated at the expected distance, 3) the difference is still recognized even they are too far from their screen. Given their response\footnote{For instance, if one answers (2) wrongly but (3) correctly, it will be categorized as sitting \textit{too far} from their screen.}, there might be a feedback message asking them to adjust their seating distance to 1.5 times their screen height. An example from the paired-comparison questions is given in Figure~\ref{fig_distance}.

\subsubsection{Training and rating}
In the training section, participants should watch the videos from the training set and rate their quality. The training set should represent the dataset under the test and cover the entire range of the rating scale. We also added a trapping clip to the training set, which instructs the participant to make a specific selection with text in the video. In case participants provide a wrong answer, they will be alerted and asked to watch the video again. Finally, the training section is shown periodically (every 60 minutes) to keep its anchoring effect following best practices in the speech quality domain \cite{naderi_open_2020}.

Within the rating section, participants rate the quality of ten video clips, one trapping, and one gold clip. The trapping and gold clip will be inserted automatically and will be used in the data cleansing step in post-processing. The video playback controller and the rating scale are presented in Figure~\ref{fig_acr}.

\section{Validation}
\label{sec:validation}
We used the VQEG HDTV datasets \cite{video2010report} to evaluate the validity and reliability of our implementation. 
The datasets contain coding only and coding plus transmission error impairments. They were created to validate objective video quality models that predicted the quality of High Definition Television (HDTV). The video materials and subjective data from experiments VQEGHD1-5 (in a laboratory) are made publicly available in the Consumer Digital Video Library\footnote{\url{https://www.cdvl.org}}. Each experiment includes 168 video clips, with 24 shared between all experiments. Each experiment includes its specific dataset where 21 HRCs were applied to 9 source videos (i.e., 144 processed sequences). We used the datasets from experiments VQEGHD3 and VQEGHD5 for which 40" (native resolution 1920x1080) and 24" (native resolution 1920x1200) displays were used in the laboratory viewing sessions by 24 test participants, respectively. The test sequences have various bit rates; 1-15 Mbps for VQEGHD3 and 2-16 Mbps for VQEGHD5. The test sequences created from two source clips in the dataset VQEGHD5 were not included in the published package, leading to 136 sequences. Results of laboratory-based subjective tests were published in \cite{video2010report} by VQEG.

To prepare the video clips for our crowdsourcing experiments, we re-encode the video sequences using FFmpeg with the H.264/AVC codec keeping the original frame rate, and resolution with a GOP of one second and Constant Rate Factor of 17\footnote{
ffmpeg -i INPUT -y -preset veryslow -keyint\_min 2 -g 24 -sc\_threshold 0 -c:v libx264 -pix\_fmt yuv420p -crf 17 OUT.mp4
}.

We conducted six crowdsourcing studies with the two above-mentioned datasets. In one of the tests, the VQEGHD5 dataset was used. In the rest of the five tests, we used the VQEGHD3 dataset on five separate days, each with unique raters, to evaluate the reproducibility of our implementation.
In all tests, participants with a minimum display resolution of 1280x720 and a refresh rate of 30Hz were allowed.

\subsection{Accuracy}
In each experiment, 10 test sequences, one gold clip, and one trapping clip were presented in one session and we aimed to collect 30 ratings per clip.
In the five experiments using the VQEGHD3 dataset an average of 71\% submissions passed the data cleansing step; the rest were not used due to the rater providing a wrong answer to the gold question, longer playback duration, failure on the second brightness check, low variance in ratings, or a wrong verification code. On average we had 21 accepted votes per test sequence (with a minimum of 15 ratings).

In the sixth experiment with the dataset VQEGHD5, 78.8\% of submissions passed the data cleansing step; the rest were not used due to similar issues. On average we had 25 accepted votes per test sequence (with a minimum of 20 ratings).

\begin{figure*}[tb]
    \centering
    \subfloat[]{\includegraphics[width=0.3\textwidth]{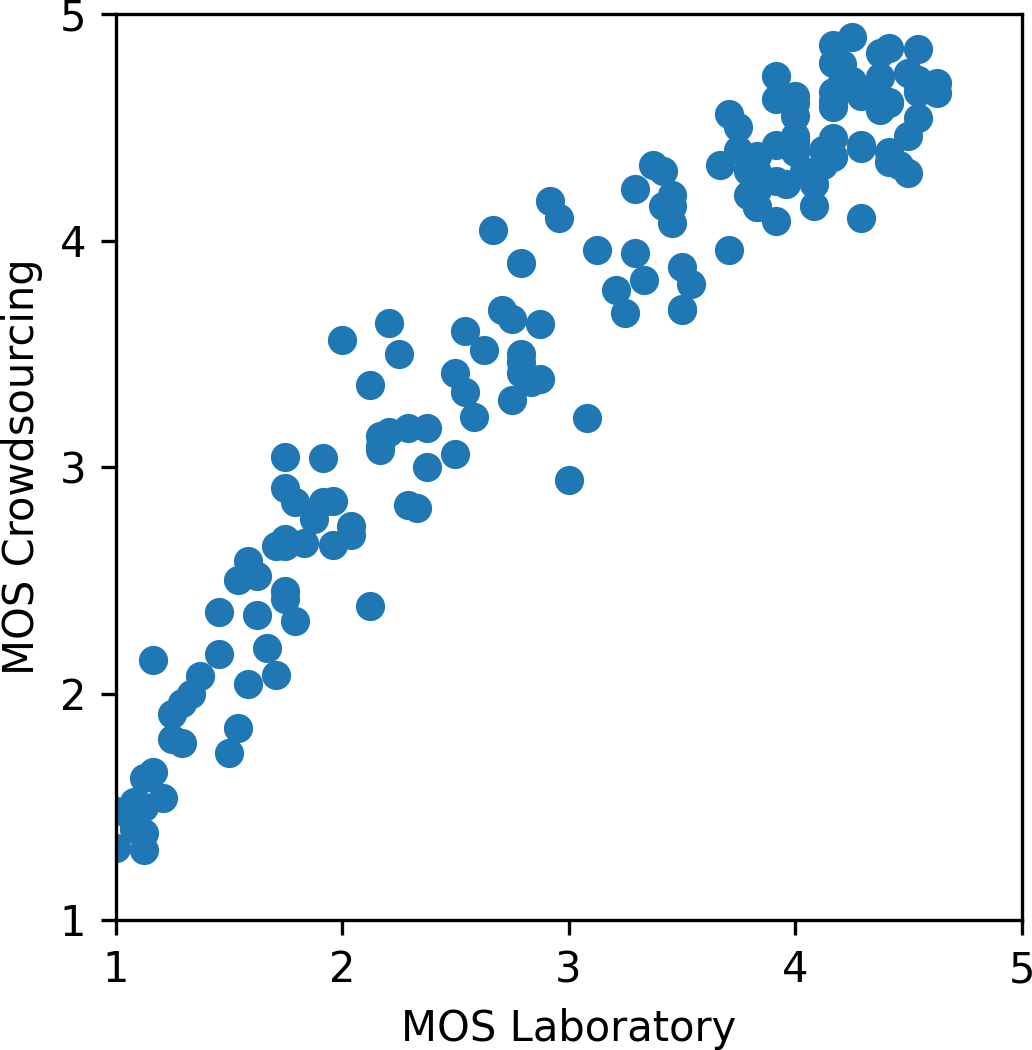}}
    \subfloat[]{\includegraphics[width=0.3\textwidth]{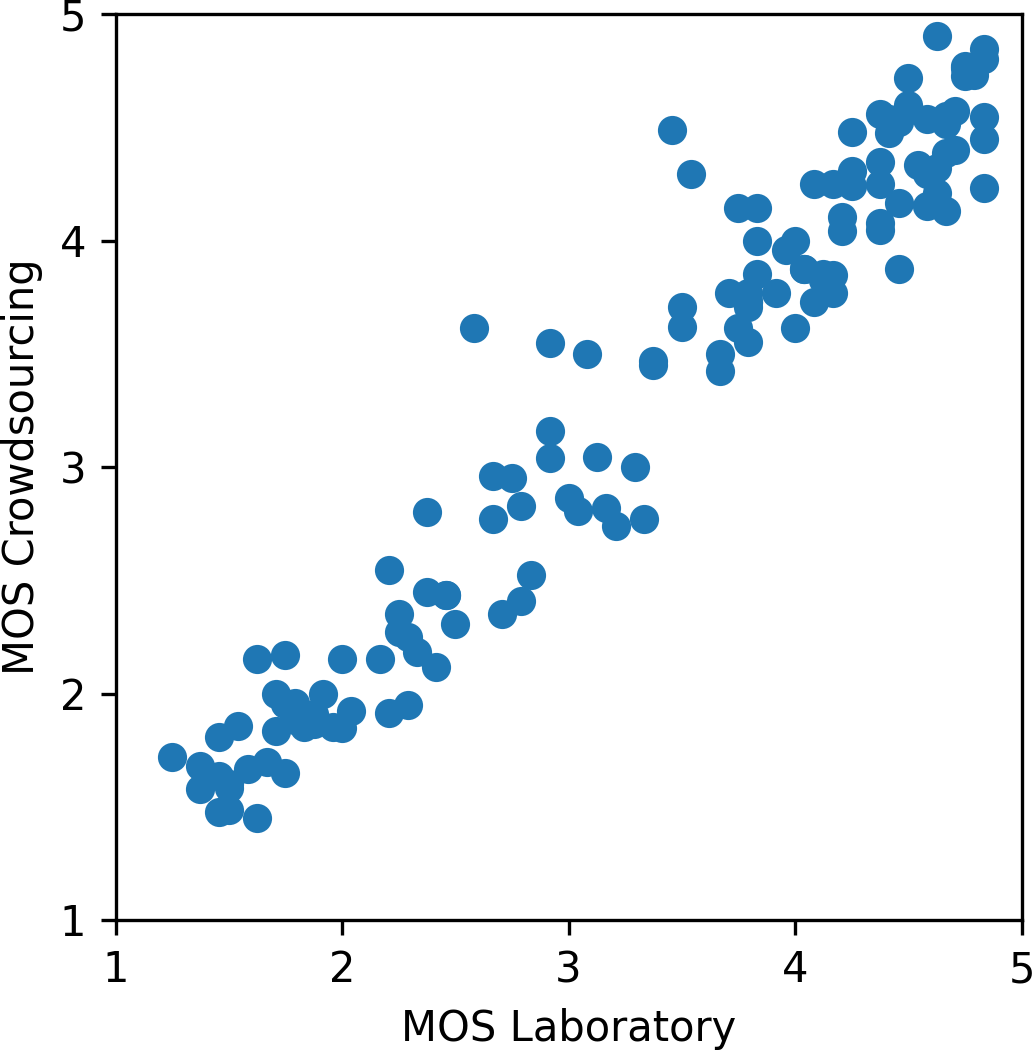}}
    \subfloat[]{\includegraphics[width=0.3\textwidth]{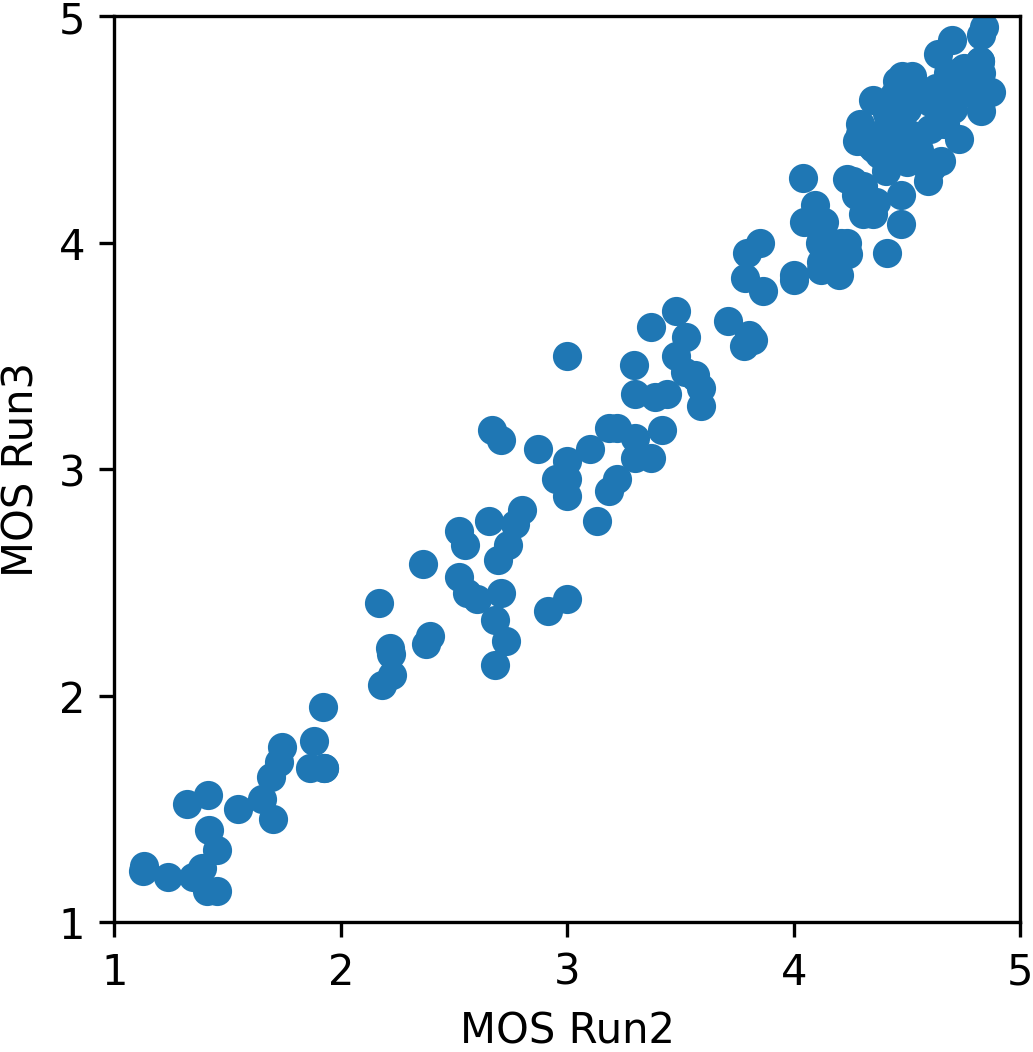}}
    \caption{Distribution of MOS values between laboratory-based and crowdsourcing-based experiments (a-b) and two crowdsourcing runs (c). (a) Dataset VQEG HD3-run1 and (b) Dataset VQEG HD5.}
    \label{fig:results:all}
\end{figure*}

The results are presented in Table~\ref{tab:validity} per test sequences and in Table~\ref{tab:validity_cond} per HRCs. We observed a strong correlation between laboratory and crowdsourcing subjective tests ($\overline{r} = 0.952$ per test sequence and $0.964$ per HRC). The distribution of MOS values from crowdsourcing and laboratory tests, before a First-Order Mapping (FOM)\footnote{FOM is typically used to removed bias and gradient between scores from two subjective tests without changing the rank-order\cite{ITU-P1401}. Here, to visualize the effect, the scatter plots are illustrated before without FOM however RMSEs are calculated after mapping.}, which is illustrated in Figure~\ref{fig:results:all}.

\begin{table}[tb]
    \caption{Comparison between laboratory and crowdsourcing experiments (sequence level).}
    \label{tab:validity} 
    \begin{center}
    \resizebox{\columnwidth}{!}{%
        \begin{tabular}{ l c c  c  c  c  c }
        \toprule
        \textbf{Dataset} &	 \multicolumn{3}{c}{\textbf{MOS}} &
         \multicolumn{3}{c}{\textbf{DMOS}} \\
        
        & {\small \textbf{PCC}}&	{\small\textbf{SPCC}}&	{\small\textbf{RMSE FOM}} & {\small \textbf{PCC}}&	{\small\textbf{SPCC}}&	{\small\textbf{RMSE FOM}} \\ 
        \midrule
        VQEG HDTV3  -run1 & 0.956 & 0.949 & 0.333 & 0.948 & 0.949 & 0.362\\
        VQEG HDTV3  -run2 & 0.964 & 0.951 & 0.302 & 0.946 & 0.939 & 0.370\\
        VQEG HDTV3  -run3 & 0.959 & 0.949 & 0.323 & 0.940 & 0.942 & 0.389\\
        VQEG HDTV3  -run4 & 0.917 & 0.913 & 0.455 & 0.904 & 0.922 & 0.489\\
        VQEG HDTV3  -run5 & 0.947 & 0.923 & 0.367 & 0.932 & 0.909 & 0.415\\
        VQEG HDTV5        &	0.970&	0.957 &	0.278 & 0.965 & 0.958 & 0.299\\ 
        \bottomrule
        \end{tabular}
    }
    \end{center}
\end{table}

\begin{table}[htbp]
    \caption{Comparison between laboratory and crowdsourcing tests in HRC level
    }
    
    \label{tab:validity_cond} 
    \begin{center}
    \resizebox{\columnwidth}{!}{%
        \begin{tabular}{ l c c  c  c }
        \toprule
        \textbf{Dataset} &	 \multicolumn{3}{c}{\textbf{MOS}} \\

        & {\small \textbf{PCC}}&	{\small\textbf{SPCC}}&	{\small\textbf{RMSE}} & {\small\textbf{RMSE FOM}}  \\ 
        \midrule
            VQEG HDTV3 -run1 &0.967	& 0.980	& 0.655	& 0.248\\
            VQEG HDTV3 -run2 &0.977	& 0.982	& 0.618	& 0.211\\
            VQEG HDTV3 -run3 &0.968	& 0.981	& 0.577	& 0.245\\
            VQEG HDTV3 -run4 &0.940	& 0.975	& 0.706	& 0.333\\
            VQEG HDTV3 -run5 &0.965	& 0.972	& 0.671	& 0.257\\
        \bottomrule
        \end{tabular}
    }
    \end{center}
\end{table}

\subsection{Reproducibility}
 On average 63 unique workers participated in each run. We observed a $PCC=0.971$ and $SRCC=0.95$ on average between the MOS values of sequences in the five different runs. The correlation coefficients between the runs are reported in Table~\ref{tab:repro}.
We also fitted a linear mixed-effects model (LMEMs) with random intercept in which test sequences and runs are considered as fixed factors and participants as a random factor. The result shows there was no significant effect of runs on the subjective ratings ($\chi^2(4)=2.691, p=0.611$).

\begin{table}[tb]
    \caption{Correlation coefficients between five runs of the VQEGHD3 dataset. Pearson correlation coefficients are on the upper triangle and Spearman's rank correlation coefficients are on the lower triangle.}
    \label{tab:repro} 
    
    \begin{center}
    \resizebox{0.8\columnwidth}{!}{%
        \begin{tabular}{ c c c  c  c  c }
        \toprule
        &\textbf{Run 1} &\textbf{Run 2} &\textbf{Run 3} &\textbf{Run 4} &\textbf{Run 5} \\ 
        \midrule
        \textbf{Run 1} &         &0.984  &0.987  &0.957   &    0.977\\
        \textbf{Run 2} &0.959    &       &0.985  &0.957   &    0.977\\
        \textbf{Run 3} &0.974    & 0.969 &       &0.952   &    0.972\\
        \textbf{Run 4} &0.943    & 0.941 & 0.942 &        &    0.956\\
        \textbf{Run 5} &0.954    &0.947  & 0.942& 0.933   &    \\
        
        \bottomrule
        \end{tabular}
    }
    \end{center}
\end{table}

\subsection{Ablation study}
We evaluated the effect of different components on the performance of the crowdsourcing test. 
In pretests, we observed that the percentage of correct answers to the brightness check increases by 19\% when participants are asked to perform screen calibration.
Furthermore, in preliminary studies, we observed that in 24.8\% of submissions, the playback duration was longer than 1.15 times the original duration of the videos. After a software update, in which videos are fully downloaded in advance, only 5.06\% of cases showed longer playback duration than expected. That change led to a 0.04\% increase in correlation coefficients to lab accuracy and a 0.1 reduction of RMSE.

We evaluated the performance of the gold clips, playback duration, brightness check, and straightliner\footnote{Straitliners select the same rating for many consecutive clips.} detection by comparing the MOS values from the laboratory test (as the ground truth) with the MOS values calculated from submissions that passed or failed in each criterion. 
We combined submissions from all five runs in the reproducibility study. The results are presented in Table~\ref{tab:performance_1}.
The correlation coefficients between the laboratory scores and the aggregated scores that passed all the criteria are significantly higher than the other groups. Similarly, the RMSEs were significantly lower. 

\begin{table}[t]
\caption{Effect of integrated filtering criteria on the validity of the crowdsourcing test.}
\label{tab:performance_1} 
\begin{center}
\resizebox{\columnwidth}{!}{%
    \begin{tabular}{l c c c c c}
    \toprule
    \textbf{Case} & \textbf{Avg.\# Ratings} & \textbf{PCC} & \textbf{SRCC} & \textbf{RMSE} & \textbf{RMSE FOM} \\
    \midrule
    All passed & 104 &  0.96 & 0.96 & 0.62 & 0.31 \\   
    Gold clips failed &  17 & 0.57 & 0.53 & 1.02 & 0.93 \\
    Playback duration failed &  8 & 0.62 &0.57 & 1.12 & 0.89 \\
    Brightness check failed&  17 & 0.89 & 0.88 & 0.84  & 0.52\\
    Straightliners  & 6 &0.29 & 0.30 & 1.53 & 1.09\\
    \bottomrule
    \multicolumn{6}{l}{All statistics are significantly different from their counterpart in \textit{All passed} at $\alpha=.05$}
    \end{tabular}%
}
\end{center}
\end{table}

The rest of the components are evaluated in separate studies which are detailed in the following sections.

\subsubsection{Viewing distance}
We conducted a crowdsourcing experiment with two settings in which participants were asked 1) to sit at 1.5 times the height of their screen and 2) to sit as far as they could but still be able to interact with their device and not be closer than three times of the height of their screen. The setup section (including the viewing distance check) was forced to be shown in all instances for both settings. Despite the original viewing distance test, no feedback to participants was provided upon their answers. Submissions that passed all the data cleansing criteria are aggregated based on their performance in the viewing distance test.
The results are presented in Table~\ref{tab:performance_vd}. The correlation coefficient between the ratings from the submissions that passed the viewing distance and the laboratory scores is significantly higher than those that failed the test ($z=4.435$, $p<.001$).

\begin{table}[t]
\caption{Effect of the viewing distance (VD) component on the validity of the crowdsourcing test.}
\label{tab:performance_vd} 
\begin{center}
\resizebox{1\columnwidth}{!}{%
    \begin{tabular}{l c c c c}
    \toprule
    \textbf{Case} & 
    \textbf{PCC} & 
    \textbf{SRCC} & 
    \textbf{RMSE} & 
    \textbf{RMSE FOM} \\
    \midrule
    VD - Passed     & 0.93 &    0.92 &  0.76 &  0.41 \\
    VD - Failed     & 0.83 &    0.78 &  0.95 & 0.62  \\
    VD - Failed (too far)    & 0.87 &    0.84 &  0.81\textsuperscript{*} &  0.58 \\
    VD - Failed (unknown category)     & 0.79 &    0.72 &  1.07 &  0.70 \\
    \bottomrule
   \multicolumn{5}{l}{\textsuperscript{*} Beside this value, all statistics are significantly
   different from their counterpart in} \\
   \multicolumn{5}{l}{ \textit{VD- passed} at $\alpha=.05$}
    \end{tabular}%
}
\end{center}
\end{table}

\subsubsection{Visual acuity test}
In a new crowdsourcing test, the task was modified so that the Visual Acuity Test (VAT) did not give any feedback on its finish, and despite its results, participants could continue to the rating section. Meanwhile, the qualification section (which contains the VAT) was included in all tasks. To increase the diversity of participants, up to five tasks per worker were allowed. We used the VQEG HD3 dataset for this test as well.
In post-processing, we grouped submissions based on their performance in the VAT (passed or failed) and the rest of the quality control criteria. 13.3\% of submissions that passed all the other quality control criteria failed in VAT. For each group, we aggregated the ratings per test sequence and compared the resulting MOS values with scores from the laboratory tests. The results are presented in Table~\ref{tab:performance_vat}. The correlation of laboratory scores to MOS values from the group that passed the VAT and all the other criteria is higher than the group that failed in VAT but passed the other criteria ($z = 1.76$, $p = 0.039$). 

\begin{table}[t]
\caption{Effect of the Visual acuity test (VAT) component on the validity of the crowdsourcing test.}
\label{tab:performance_vat} 
\begin{center}
\resizebox{1\columnwidth}{!}{%
    \begin{tabular}{l c c c c}
    \toprule
    \textbf{Case} & 
    \textbf{PCC} & 
    \textbf{SRCC} & 
    \textbf{RMSE} & 
    \textbf{RMSE FOM} \\
    \midrule
    VAT Passed \& All criteria passed  & 0.91 &    0.90 &  0.88 &  0.47 \\
     VAT Failed \& All criteria passed & 0.87 &    0.89 &  0.96 &  0.59  \\
     \midrule
    VAT Passed \textsuperscript{*} & 0.91 &    0.90 &  0.97 &  0.48 \\
    VAT Failed \textsuperscript{*} & 0.63 &    0.59 &  1.41 &  0.89 \\
    \bottomrule
   \multicolumn{5}{l}{\textsuperscript{*}No other criterion was considered.} 
    \end{tabular}%
}
\end{center}
\end{table}

\begin{figure}[htb]
    \centering
    \subfloat[]{\includegraphics[width=0.8\columnwidth]{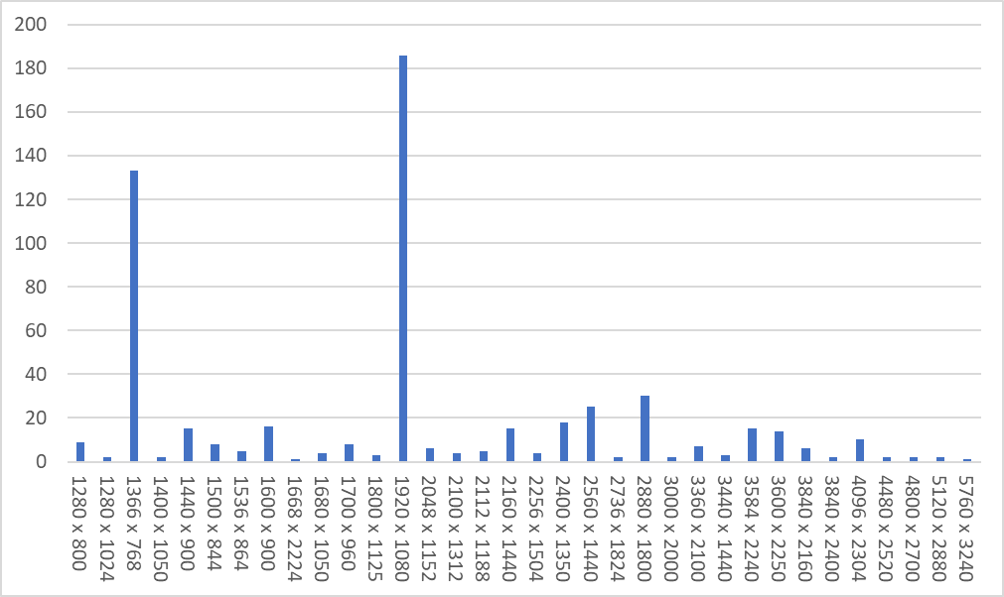}%
    \label{fig:monitor_size}
    }
    
    \subfloat[]{\includegraphics[width=0.8\columnwidth]{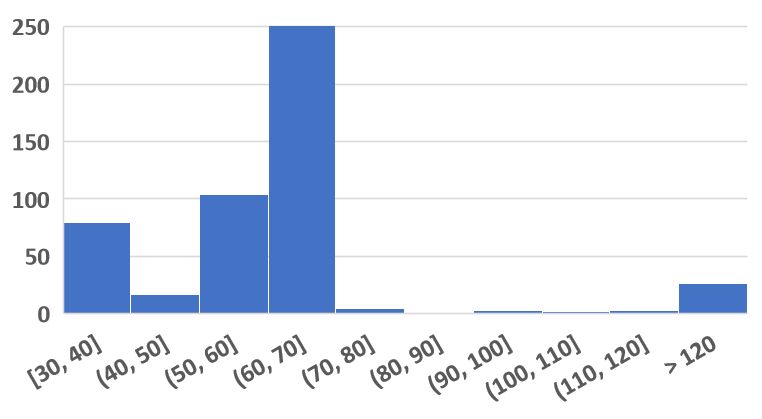}%
    \label{fig:monitor_refresh}
    }
    
    \subfloat[]{\includegraphics[width=0.8\columnwidth]{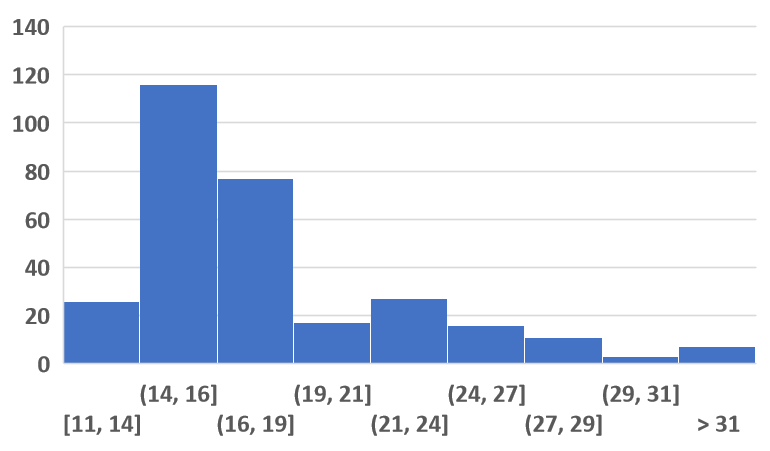}%
    \label{fig:monitor_inch}
    }
    \caption{Properties of crowd workers' display. (a) Resolution, (b) Estimated refresh rate (Hz), (c) Estimated display size (inch).}
\end{figure}

\subsubsection{Display calibration and trapping clips}
To validate the display calibration and trapping question, we conducted three new crowdsourcing experiments. In the first experiment, we modified the test and left out the display calibration task (and the brightness/light check). The trapping clips from the training and rating sections were left out in the second experiment. In the last one, the complete test was used. In all experiments, the participants rated the VQEG HD3 dataset, and we aimed to collect 30 votes per clip.
The results are presented in Table~\ref{tab:performance_trap_calib}.
The correlation coefficients between laboratory and crowdsourcing scores were significantly higher when the complete test was used. In the same case, the RMSE after a first-order mapping was significantly lower. We also conducted multiple Welch’s t-tests (one per test sequence) with Bonferroni correction to determine which sequences are rated significantly different in the crowdsourcing tests compared to the laboratory. In the case of \textit{complete test}, only 16\% of the sequences were rated significantly different than the laboratory results. Half of those cases belong to HRC 16 and 17, in which videos were coded with 1.0 and 1.2 Mbps bitrates. We anticipate the difference is due to the fact that the laboratory test was conducted using a 40'' screen, whereas crowd workers used their PC and laptop screens (typically 11 to 24''). The distribution of display resolution and estimated display refresh rate of crowd workers who participated in our studies are presented in Figure~\ref{fig:monitor_size} and  Figure~\ref{fig:monitor_refresh}, respectively.

%

\begin{table}[t]
\caption{Effect of calibration task and trapping clips.}
\label{tab:performance_trap_calib} 
\begin{center}
\resizebox{1\columnwidth}{!}{%
    \begin{tabular}{l c c c c c}
    \toprule
    \textbf{Case} & 
    \textbf{PCC} & 
    \textbf{SRCC} & 
    \textbf{RMSE} & 
    \textbf{RMSE FOM} &
    \textbf{Sig. diff. to lab}\\
    \midrule
    Complete test & 0.95 &    0.95 &  0.72 &  0.34 & 16\%  \\
    No Calibration & 0.91 & 0.89 &  0.80\textsuperscript{*}&  0.47 & 23\%  \\
    No Trapping clip & 0.92 &   0.92 &  0.88 &  0.46 & 29\%  \\
    \bottomrule
    \multicolumn{6}{l}{\textsuperscript{*} Beside this value, all statistics are significantly
   different compared to the \textit{Complete test}} \\
   \multicolumn{6}{l}{ at $\alpha=.05$} 
    \end{tabular}%
}
\end{center}
\end{table}

\subsection{Number of ratings}
\begin{figure}
    \centering
    \subfloat[]{\includegraphics[width=0.8\columnwidth]{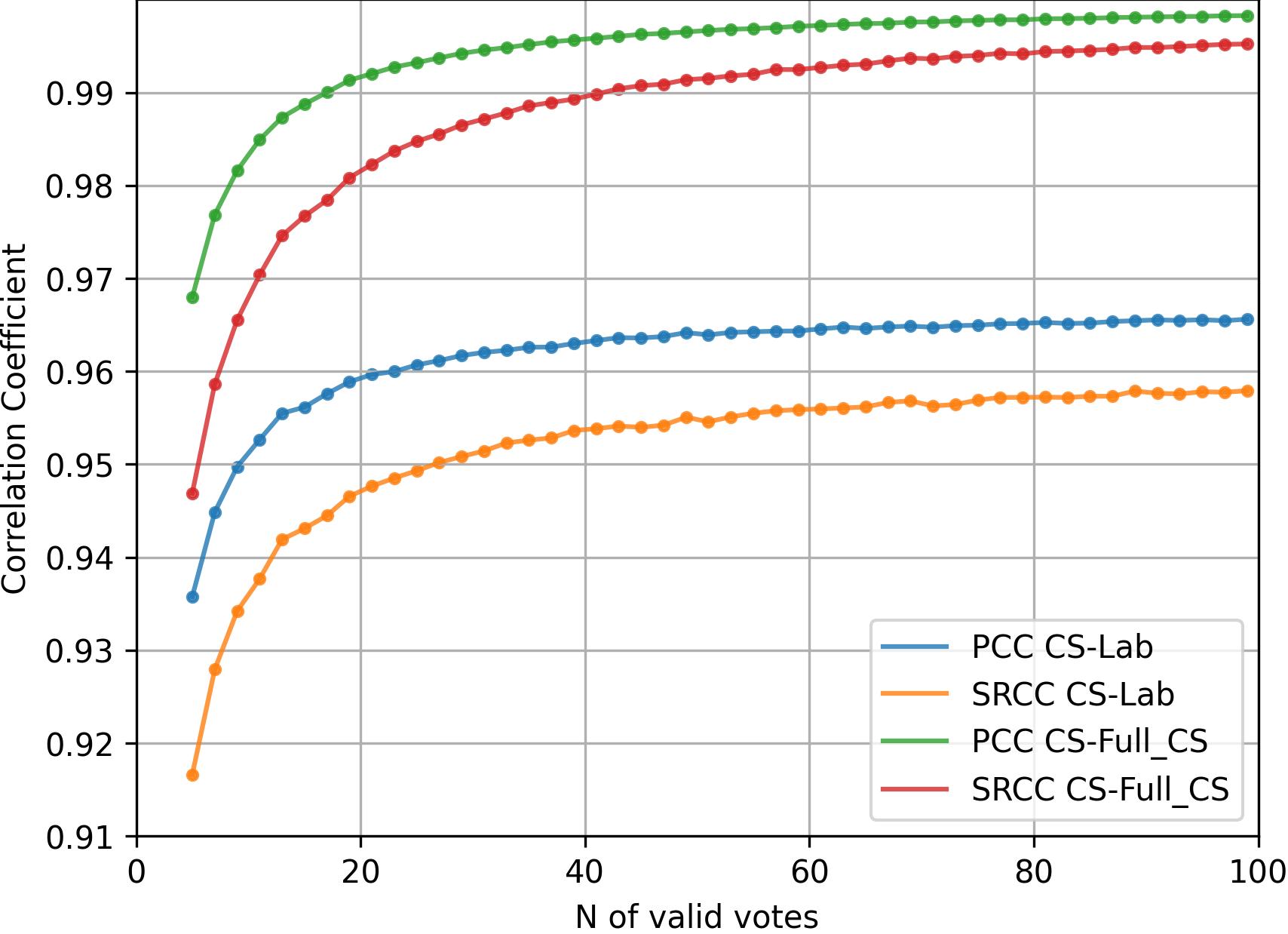}%
    \label{fig:coef:change}}

    \subfloat[]{\includegraphics[width=0.8\columnwidth]{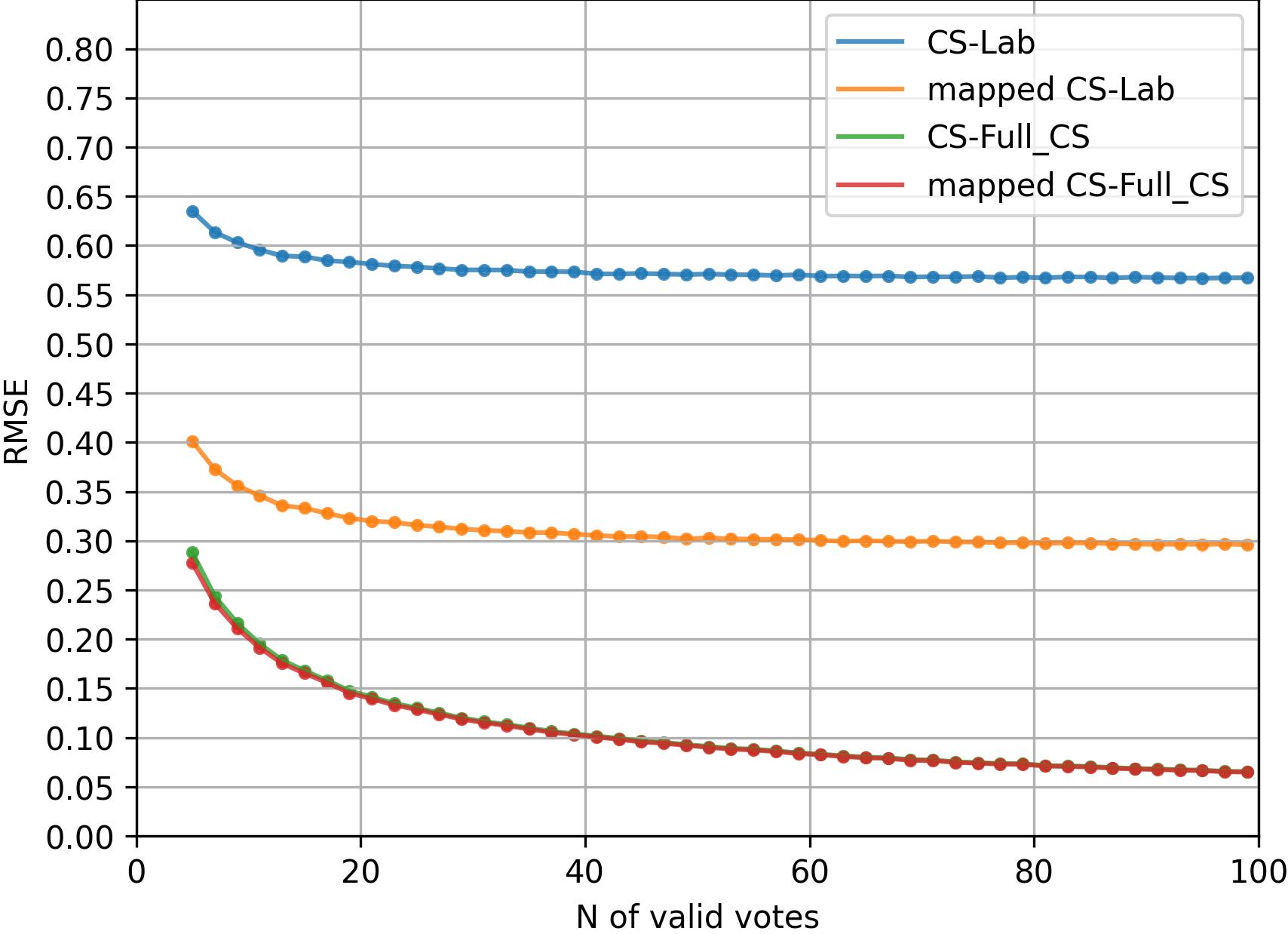}%
    \label{fig:rmse:change}}
    
    \caption{Changes of different statistics by increasing the number of votes used for calculating MOS, (a) Correlation coefficients, (b) RMSE.}
    \label{fig:n_votes}
\end{figure}

We combined all the crowdsourcing studies in which the dataset VQEG HDTV3 and the complete set of quality control mechanisms were used.
We conducted a bootstrapping simulation study to determine how the accuracy of MOS scores changes depending on the number of valid votes per sequence.
In the bootstrapping process, we randomly selected N votes per sequence with replacement and calculated the MOS values per sequence and consequently correlation and RMSE between those MOS values and MOS values from Lab or MOS values from crowdsourcing when all valid votes are used. We repeated this process 200 times per the number of votes (N) and calculated the mean and 95\% confidence interval for all statistics.
Figure~\ref{fig:coef:change} illustrated changes in correlation coefficients and Figure~\ref{fig:rmse:change} changes in RMSE by increasing the valid number of votes used for calculating MOS.
The results show that both coefficients and RMSE plots saturated on about 40 votes.


\section{Test Methods}
We compared the quality of compressed videos submitted to the CVPR 2022 CLIC challenge using the ACR, DCR, and CCR test methods.
Each of the 12 teams compressed the 30 videos in the test set targeting 1 Mbps and 0.1 Mbps average bitrates during the challenge, creating an overall 27 sets of clips including the original sequences and two sets encoded by H.264 and AV1.
On average, we have collected 21.6 valid votes per sequence in the ACR test, 12.1 in the DCR test (on a 9-point discrete scale), and 15.2 in the CCR test. As the original sequences were included in the test set, we also calculated the Differential MOS which is typically the result of the ACR with Hidden References (ACR-HR) test method.  
Finally, we aggregated the ratings on the sequence (i.e., clip) and also on the model (i.e., submission of a team for 0.1 MBps or 1 MBps task) level. Table~\ref{tab:test_methods} reports the correlation coefficients between aggregated ratings. 
The highest correlation was observed between the ratings from ACR and ACR-HR (single stimulus methods) and between the DCR and CCR methods (double-stimulus methods). Correlations to the most used objective measures are also reported in Table~\ref{tab:test_methods_obj}, which shows a poor correlation with subjective scores.

The \P910 recommends using double-stimulus test methods with explicit reference when the transparency or fidelity of the underlying process or transmission should be evaluated or when high-quality systems are under test. These methods lead to longer test sessions or more participants and consequently higher costs compared to the single-stimulus approach (about $\times2$). We also observed a large difference in the quality ranking of sequences according to single- or double-stimulus tests when they contain 1) imperfect source sequence (e.g., low image quality, or slow motioned), 2) specific impairment due to encoding process (e.g., color changes or blur effect on non-saliency area), 3) source sequence with bokeh.  

\begin{table}[htbp]
    \caption{Pearson correlations (upper-triangle) and Spearman's rank correlations (lower-triangle) between ratings collected via different test methods, aggregated in model and clip level.}
    
    \label{tab:test_methods} 
    \begin{center}
    \resizebox{\columnwidth}{!}{%
        \begin{tabular}{ c  c c c c  c   c c cc}
        \toprule
         \textbf{Test}&	 \multicolumn{4}{c}{\textbf{Model level}}& &  \multicolumn{4}{c}{\textbf{Clip level}}\\
        
        \textbf{method} & {\small \textbf{ACR}} & {\small \textbf{ACR-HR}}&	{\small\textbf{DCR}}&	{\small\textbf{CCR}} & & {\small \textbf{ACR}}& {\small \textbf{ACR-HR}} &	{\small\textbf{DCR}}&	{\small\textbf{CCR}} \\ 
        \midrule
            {\small\textbf{ACR}} &   -  & 0.999  & 0.981 & 0.963 & &   -     & 0.975 &    0.948 & 0.920\\
            
            {\small\textbf{ACR-HR}} & 0.999 &  -    & 0.983 & 0.966 & &  0.964 & -     &    0.939 & 0.912\\
            
            {\small\textbf{DCR}} & 0.976 & 0.981 &  -    & 0.990 & & 0.948 & 0.935  &   -      & 0.962\\
            {\small\textbf{CCR}} & 0.987 & 0.989& 0.981  &  -     & & 0.936 &0.928  & 0.960    &   -   \\
        \bottomrule
        \end{tabular}
    }
    \end{center}
\end{table}

\begin{table}[htbp]
    \caption{Correlation between ratings collected via different subjective test methods and prediction of objective measures in clip level.}
    
    \label{tab:test_methods_obj} 
    \begin{center}
    \resizebox{\columnwidth}{!}{%
        \begin{tabular}{ c  c c c   c   c c c}
        \toprule
         \textbf{Test}&	 \multicolumn{3}{c}{\textbf{Pearson correlation}}& &  \multicolumn{3}{c}{\textbf{Spearman's rank correlation}}\\
        
        \textbf{method} & {\small \textbf{PSNR}} & {\small \textbf{MS-SSIM}}&	{\small\textbf{VMAF}} & & {\small \textbf{PSNR}}& {\small \textbf{MS-SSIM}} &	{\small\textbf{VMAF}}	 \\ 
        \midrule
            {\small\textbf{ACR}} &  0.691  & 0.757  & 0.914  &  &   0.691     &  0.830  &   0.896  \\
            {\small\textbf{ACR-HR}} & 0.725 & 0.773 & 0.919  &  &   0.731     &  0.855  &   0.903  \\
            {\small\textbf{DCR}} & 0.735 & 0.722 & 0.915     &  &   0.751     &  0.861  &   0.914   \\
            {\small\textbf{CCR}} & 0.706 & 0.688 & 0.893     &  &   0.713     &  0.843  &   0.910   \\
        \bottomrule
        \end{tabular}
    }
    \end{center}
\end{table}

\section{Discussion and Conclusions}

We have created the first crowdsourced implementation of \P910 that is shown to be as accurate as a lab study and is highly reproducible. In a set of experiments, we showed that each test component significantly improves the validity of the collected ratings. 

It should be noted that the impact of the test components depends on the reliability of crowd workers (and their environment and device) and the test sequences under study. If a dedicated group of workers performs the test in a suitable environment using proper devices, those components might look redundant. However, they strongly increase the validity of collected scores and robustness of the test method despite the group of test participants and consequently make large-scale data collection possible. 
Furthermore, we let the crowd workers watch the videos as often as they want, in contrast to the strict one-time play-back policy and limited time for casting votes typically used in laboratory tests. We chose this approach to compensate for the uncontrolled nature of the crowd worker's environment, allowing participants to watch the videos again in case of unexpected interruptions in their flow. Meanwhile, the framework logs the number of times a video sequence is played back. Within our experiments, only 6.3\% of videos were played back more than once. 

We plan to include the simultaneous presentation of sequence pairs for future work and include \P910 Annex E: An advanced data analysis technique for tests under challenging conditions. In addition, this framework can be easily extended to measure image quality assessment, which is also missing an open-source crowdsourcing implementation with rater, environment, and device qualifications as well as accuracy and reproducibility studies. 

\label{sec:conclusions}
\bibliographystyle{IEEEtran}
\bibliography{IC3-AI,other}


\vfill

\end{document}